\documentclass[10pt,conference,a4paper]{IEEEtran}
\usepackage{times,amsmath,epsfig,epstopdf, tabularx, array}

\title{An Enhanced Multi-Pager Environment Support \\ for Second Generation Microkernels}

\author
{
	{Yauhen Klimiankou}

	\vspace{1.6mm}\\
	\fontsize{10}{10}\selectfont\itshape
	
	Department of Information Technologies Software, Belarusian State University of Informatics and Radioelectronics  \\
	Minsk 220 113, Belarus\\

	\fontsize{9}{9}\selectfont\ttfamily\upshape

	klimenkov@bsuir.by | Evgeny.Klimenkov@gmail.com\\

	\vspace{1.2mm}\\
	\fontsize{10}{10}\selectfont\rmfamily\itshape
}

\begin{document}

\maketitle

\begin{abstract} 
	The main objective of this paper is to present a mechanism of enhanced paging support for the second generation microkernels in the form of explicit support of multi-pager environment for the tasks running in the system. 
	Proposed mechanism is based on the intra-kernel high granularity pagers assignments per virtual address space, which allow efficient and simple dispatching of page faults to the appropriate pagers.
	The paging is one of the major features of the virtual memory, which is extensively used by advanced operating systems to provide an illusion of elastic memory.
	Original and present second generation microkernels provide only limited, inflexible and unnatural support for paging. 
	Furthermore, facilities provided by current solutions for multi-pager support on the runtime level introduce an overhead in terms of mode switches and thread context switches which can be significantly reduced. 
	Limited paging support limits the attractiveness of the second generation microkernel based systems use in real-life applications, in which processes usually have concurrent servicing of multiple paging servers. 
	The purpose of this paper is to present a facilities for the efficient and flexible support of multi-pager environments for the second generation microkernels.  
	A comparison of the proposed solution to the present architecture L4 + L4Re has been made and overhead of the page fault handling critical path has been evaluated.
	Proposed solution is simple enough and provides a natural and flexible support of multi-pager environments for second generation microkernels in efficient way. 
	It introduces a third less overhead in terms of the mode switches and thread context switches in comparison to the present L4 + L4Re solution implemented in the Fiasco.OC.
\end{abstract}


\section{INTRODUCTION}
	This paper describes the design of the fine-grained multi-pager environment support facilities for the second generation microkernels that allows processes running in the system to be serviced by multiple pager servers concurrently and in efficient way.	
	The proposed approach describes modifications of the virtual memory management subsystem of the second generation microkernels.

	The user mode page fault handling was originally proposed by the Mach project \cite{Accetta86mach:a} \cite{conf/mach/SechrestP91}. 
	The general idea of the proposed approach is to allow the page fault handling servers to be running as a separate user mode processes.
	Safe and efficient memory management is a fundamental requirement for a microkernel. 
	Due to this, for example, substantial memory overhead imposed by originary recursive address space construction can be considered as an enough drawback to reject this memory management approach \cite{nicta_6930}.
	Traditional approach for paging support in the L4 like microkernels family is limited and not efficient for multi-pager environments, which are ordinary for the advanced real-world systems and applications.
	Insufficient support of such environments limits the attractiveness of second generation microkernel based operating systems for the real system implementations.
	Furthermore this insufficient support of paging looks inadequate to the actual state of facts, because experience gained from the ubiquitous monolithic kernels shows that the typical application are likely serviced by multiple memory management subsystems concurrently.
	For example, the typical processes in Windows \cite{Russinovich:2009:WII:1717352} and Linux \cite{Bovet:2005:ULK:1077084} environment concurrently get the next services: automatic stack expanding/reducing, dynamically  loadable modules management, anonymous memory management, unswappable memory management, shared memory management, file mappings to memory etc.
	Furthermore some specific applications can wish to use special purpose memory management facilities along with ordinary memory management services.
	For example they can wish to use SMARTMAP-like \cite{Brightwell} memory management for performance benefits or InkTag \cite{hofmann13asplos} and Gateway \cite{confndssSrivastavaG11} -like memory management features to achieve additional security and reliability guarantees.
	
	Nevertheless	 of the extensive use of the multi-pager environments in typical applications present second generation microkernels have a limited support of it.
	The original approach taken by the L4 microkernels family \cite{Liedtke95MicrokernelConstruction} is an optional assignment of exactly one pager for each task running in the system. 
	This pager is responsible for handling of all page faults generated by the tasks to which it is assigned as a pager.
	The same approach is kept in the descendant kernels like \cite{Klein:2009:SFV:1629575.1629596}.
	Present version of the L4 (Fiasco.OC) \cite{FiascoOS} provides a tricky support of multi-pager environments through introduction of the additional level of indirection - region mapper. 
	This solution has been done on the level of runtime environment system L4Re \cite{L4Re} developed specially for the L4 microkernel in the Technical University of Dresden.
	In other words, this solution is an attempt to overcome limitations imposed by the single-pager kernel design on the level of runtime environment instead of changing kernel itself, despite the fact that the kernel is a natural location for multi-pager environment support and that this support can be implemented in the kernel in efficient way and only with negligible violation of the minimality principle.
	
	Region mapper is an additional layer of indirection in page fault handling introduced by L4Re for providing a multi-pager environment for applications running in the context of this runtime environment.
	According to this solution, each process running in L4Re environment has a special thread running inside it, which plays a role of pager for all other threads running in the same process (threads that share the same virtual address space).
	This special thread of L4Re-based process is called a region mapper.
	Its main responsibility is to manage a virtual address space layout through virtual memory management and page fault handling. 
	Region mapper do this by managing a special table which tracks which region of the virtual address space is serviced by which memory manager. 
	By using this table region mapper is able to route the page faults generated by process threads to the appropriate external memory manager server.
	As a result, the typical page fault handling process goes through next steps:
	\begin{enumerate}
		\item Page fault is generated by a thread. 
		\item CPU interrupts the faulted thread and gives control to the L4 microkernel.
		\item L4 looks at the faulted thread to identify, which thread is a pager of it (it is a region mapper for L4Re based processes).
		\item L4 suspends the faulted thread and sends page fault message for the region mapper.
		\item The region mapper looks into the table by using the fault address to find out a thread which is responsible for the faulting address as a real pager.
		\item The region mapper reflects the page fault message to the real pager.
		\item The real pager takes an actual actions to resolve the page fault reason and restart the faulted thread.
	\end{enumerate}
	
	Region manager relies to the generic abstraction of memory mapping which is called dataspace, that was initially introduced as a part of SawMill VM framework \cite{Aron_PJLED_01}. 
	Dataspace is a generic source of resources capable to be mapped as a continuous memory region to the virtual address space like anonymous memory region, memory mapped file or device, etc., and provides only generic memory management functionality.
	Dataspace is a capability-protected interface implemented by L4Re but its actual implementation is provided by external thread called dataspace manager, which is in charge of the dataspace layout and its content. 
	As a result multiple dataspaces with different implementations managed by different servers can coexist in the system concurrently.
	Besides handling page faults generated by threads attached to it, region mapper is also responsible for maintaining layout of the virtual address space which it services. 
	That means that it is responsible for inserting and removing of the dataspaces to/from virtual address space.
	And due to this it is capable to add/remove appropriate entries to/from the mapping table mentioned above to maintain it in actual and consistent state.
	The actual virtual address space region represented by dataspace is populated by pages through page faults reflected by dataspace manager.
	
	The described existing mechanism for providing the multi-pager environment in the L4 family of second generation microkernels is complex and inefficient, because it involves multiple context switches for such typical tasks as page fault handling.
	We would like to propose more natural, simple and efficient way of multi-pager environment support for the second generation microkernels, which moves this support implementation from the runtime environment layer into the kernel.
	Proposed approach introduces less processor time overhead and memory overhead with only negligible violation of the minimality principle, which can be advocated by the same arguments which are applied for intra-kernel scheduling policies implementation.

	\section{VIRTUAL MEMORY AND \\ PAGE FAULT HANDLING}

	Emergence of the virtual memory technology made a great impact on the whole future computer systems development.
	The two most significant ideas behind virtual memory are:
	\begin{enumerate}
		\item Arbitrary mappings between hardware memory layout and memory layout observed by applications.
		\item Transparent changing of mappings between hardware memory and memory layout observed by applications.
	\end{enumerate}
	Both ideas rely to the explicit support of the virtual memory by underlaying CPU architecture. 
	Significance of the virtual memory technology can be stressed by the fact that CPU uses specially dedicated block called MMU for its support. 
	In the following discussion we will focus on the second major idea of the virtual memory which is widespread called paging.
	
	From the paging point of view, typical system can be split into two domains: memory resource providers and memory resource consumers, relationships between which are mediated by CPU.
	This mediation comes in two forms: present flag in page table entry and page fault exception.
	The first one allows to mark virtual memory pages as stubs, which haven't any actual resources assigned. 
	And the second one implements a way according to which the memory resource manager can be notified about attempt to access stub virtual memory page.
	Both this features together provide a channel of implicit communication between memory resource provider and memory resource consumer, which allows transparent dynamic memory management, which memory consumer don't need to take care about.
	Memory manager is able to silently get back memory allocated to consumer earlier or allocate some additional memory to it. 
	In the same time it can be silently called by consumer in the case when it requires memory that was got back by memory manager before.

	Such reasoning and understanding of the paging in context of the virtual memory lead us to a number of conceptual conclusions about paging nature.

	 \subsection{There is only one manager per unit of physical address space represented resources}
	When multiple memory managers can coexist in the system concurrently, each of them must manage its own resources, and each resource unit in the system must be managed by exactly one memory manager.
	The hierarchical memory managers chain implemented in L4 microkernels in which all chain entries manage the same region of memory is unnatural, overcomplicated. 
	Furthermore the fact that the same thread can play role of both memory manager and memory consumer for the same memory unit completely violates original virtual memory concept.
	
	\subsection{Relationships between memory consumer and memory manager goes through a virtual address space region which consumer trust to manage to a specific manager}
	Primary communication channel between memory manager and memory consumer in virtual memory system is an implicit communication channel going through CPU with protocol which allows manager to silently give and return memory to/from consumer virtual address space and allows consumer to silently request resources through page fault exceptions.
	Memory manager is trusted entity for the memory consumer by default, because it preserves access to all resources which it provides to the consumer.
	But besides trust provided to the manager in regard to access to the data stored in memory, consumer must provide it a trust of virtual address space management.
	The only consumer responsibility is to choose to which memory manager it trusts and which region of its virtual address space.

	\subsection{Memory manager and page fault handler is a single entity}
	Virtual memory model assumes that the actions that are taken in reply to the page fault exception is targeted to provide resources requested by exception trigger and restart the trigger thread execution.
	Due to the fact that the resource providing is a responsibility of memory manager, page fault handling is its natural responsibility too.
	There is no big sense to distinguish memory manager and page fault handler as two different entities.

	\subsection {Multi-pager environment is natural for advanced operating systems}
	During long history of the virtual memory based operating systems a lot of ways of virtual memory usage have been demonstrated.
	Examples of these ways includes swappable and unswappable memory allocation, memory-mapped files implementation, IO devices access management, security and process isolation, shared memory management, intelligent DLL management etc.
	Multi-pager environment is commonly supported in the widespread industrial OS like Windows and Linux and this support is extensively used by advanced applications.

	\section{DESIGN AND IMPLEMENTATION OF MULTI-PAGER ENVIRONMENT SUPPORT}

	\subsection{Virtual address space management}

\begin{figure}[t]
		\hfil \includegraphics[width=5cm]{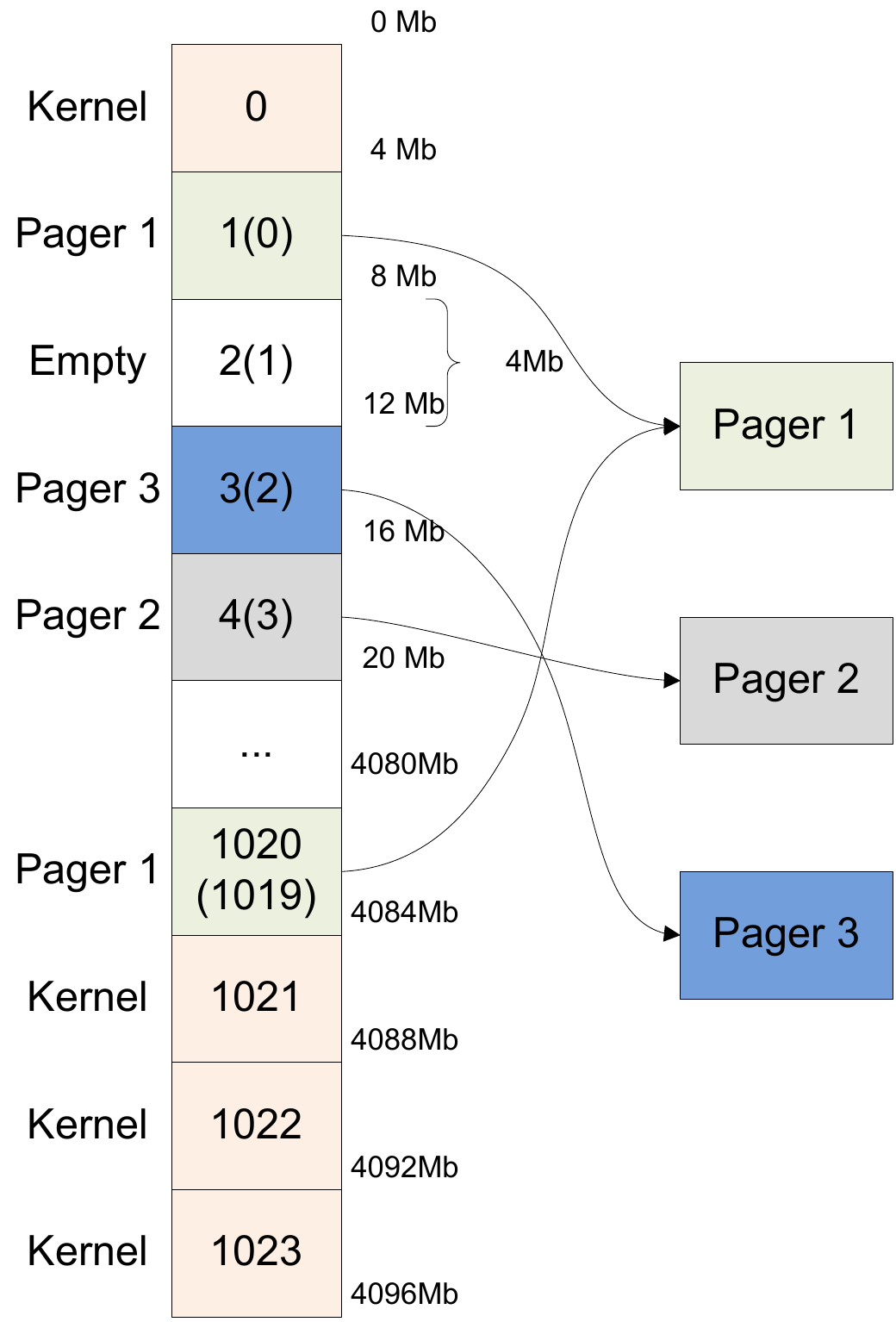} \hfil 
		\caption{Page Faults Dispatching}
		\label{fig5}
	\end{figure}

	Proposed model of multi-pager environment support introduces a fundamental abstraction - user space region.
	User space region is a fixed size continuous part of user part of virtual address space.
	User space part of virtual address space is split into a fixed number of regions, each of which contains a fixed number of pages.
	In our experimental system we have 1020 regions per virtual address space and 1024 pages per region.
	Due to this our model shown in figure \ref{fig5} resembles memory management architecture of Intel x86 \cite{Intel:2000:IIAc}, where region can be considered as an counterpart of the directory and represents 4 Mb window of virtual address space.

	User space region is a fundamental unit of virtual address space management granularity.
	Each user space region can be managed by independent manager.
	Region manager have rights to map and unmap resources owned by it into any place in the managed region without any restrictions.
	As a result it can do this completely transparently to any thread running in the context of virtual address space witch contains that region.

	Region manager plays both roles: memory manager and pager for the regions assigned to it.
	Due to this on the one hand all page faults occurred in the region are transparently routed for handling to the region manager assigned to it, and on the other hand that region manager can transparently reply to the page fault by mapping resources requested by it into the appropriate place of the region affected.

	Threads running the context of virtual address space are responsible only for assigning region managers for each particular region of its virtual address space. 
	By assigning the manager for the region thread provides to it trust of this region management and is unable to control particular mappings and unmapping operation.
	Due to the fact there are multiple regions in the same virtual address space and managers are assigned to them independently, in result proposed model represents a natural multi-pager environment with good enough management granularity.

	Implementation of the proposed model introduces memory overhead in 4Kb per virtual address space. 
	Kernel incorporates regions table into the virtual address space abstraction implementation. 
	Region table is implemented as a memory page which contains an array of thread ids of region managers. 
	To each user space region with sequential number N corresponds the regions table entry with the same sequential number.
	Regions table itself is mapped into the kernel part of the virtual address space which is located on a fixed address.
	As a result during page fault kernel can easily find the region table itself and identify the region manager responsible for the region to which fault address belongs, and to which kernel will send a page fault notification message.
	User space region id corresponding to a virtual address belonging to a user space part of virtual address space can be easily found using the next formula:
	\begin{equation}
		RID = (A_v - B_{us}) / RS,
	\end{equation}
	where RID is a region sequential number, A$_v$ is a virtual address to which this RID corresponds, B$_{us}$ is a base address of user space part of virtual address space and finally RS is a size of region. 
	Note that the B$_{us}$ and RS is a constants defined by the system design.
	Note also, that if RS is represented by value which is equal to power of two (which is a case of our implementation) the costly division operation can be replaced by cheap bit shift operation.
	
	\subsection{Page fault handling}

	Exceptions are a natural class of system events about which kernel must take care.
	In accordance with spirit of second generation microkernel design the actual work of the exception handling must be pushed out into user space and kernel must only dispatch the exception handling activities provided by external user mode servers.
	Due to this our experimental kernel doesn't distinguish exceptions of different types and handles them in the uniform way.
	
\begin{figure}[t]
		\hfil \includegraphics[width=\linewidth]{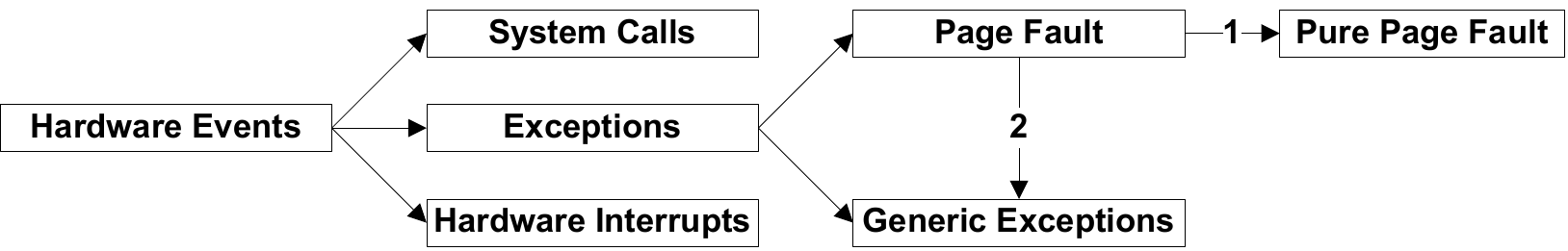} \hfil 
		\caption{Page fault classification}
		\label{fig4}
	\end{figure}

	Despite the fact, that in proposed model kernel doesn't distinguish different types of exceptions and deals with all them in uniform way, the page fault exception is considered by it as a very special type of exceptions as shown in figure \ref{fig4}.
	For page faults kernel provides an additional zero level of handling and skips all other levels of handling in the case of success on that zero level.
	
	At zero-level kernel next distinguishes two types of page faults: pure page faults and general protection page faults. 
	The first ones are faults for addresses belonging to the user part of virtual address space and generally eligible from the protection point of view.
	The second ones are an faults for addresses outside of the user part of virtual address space.
	The kernel takes special handling only for the pure page faults and consider the another page faults as a general protection faults which are an example of the generic exceptions that must be handled in the ordinary way.
	By this kernel can separate out faults that clearly aren't related to the paging and represent a clear protection violation attempt like null pointer dereferencing or attempt to access kernel code or data.

	Not all pure page faults are serviced by appropriate pagers. 
	There are two exceptional cases: 
	\begin{enumerate}
		\item Faulted virtual address belongs to the virtual address space region which has not pager assigned.
		\item Faulted virtual address belongs to the virtual address space region which has pager assigned but the assigned pager didn't accept the servicing of that region.
	\end{enumerate}
	
	The first exceptional case is a result of multi-pager environment support. 
	There is no single pager assigned to the thread, which is responsible for handling all page faults triggered by this thread despite the nature of page fault. 
	Instead there is single virtual address space split into multiple regions, each of which can have pager assigned. 
	As a result the virtual address space can be sparsely populated space, some regions of which are assigned to the pagers, and the rest have not any pagers assigned. 
	Page fault triggered in reply to attempt to access the second ones are considered as a general protection faults.

	The relationships between memory provider task (pager) and memory user task are based on the contract.
	To establish this relationships agreements of both sides must be received by the kernel.
	Memory user task provides this agreement by explicitly assigning the specified pager task to a particular virtual address space region.
	Memory provider task provides this agreement implicitly by taking memory management action on the region assigned to it.
	Furthermore pager can revoke its agreement by removing last piece of memory from the particular region with REVOKE\_AGREEMENT flag specified in the system call.
	As a result pager task can protect itself from the malicious or misbehaving memory user task which too frequently generates page faults and by this performs DoS attack on the pager.
	But from the other hand this feature introduces the second exceptional case on which the page fault is generated on the region which isn't accepted by the pager. 
	This case is also considered by the kernel as a general protection page fault.

	The last check which can take place before invoking an appropriate pager is checking of the actual not presence of the appropriate memory page. 
	On the x86 platform it can be done by checking PAGE PRESENT flag in the page table entry denoting page on which page fault occurred.
	This last check introduces negligible overhead, because according to our memory management subsystem implementation, kernel always reads page table entries during page fault handling to get 31-bit pager defined marker from the page table entry corresponding to the faulted page.
	This check can have sense, because there are multiple threads running in the same virtual address space allowed, which can potentially fault in the same page in very little period of time. 
	As a result the situations are possible, in which the faulted page can be restored by the pager between page fault and actual pager invoking. 
	
\begin{figure*}[ht]
		\centering
		
		\vspace*{\fill}

		\hfil \includegraphics[width=15cm]{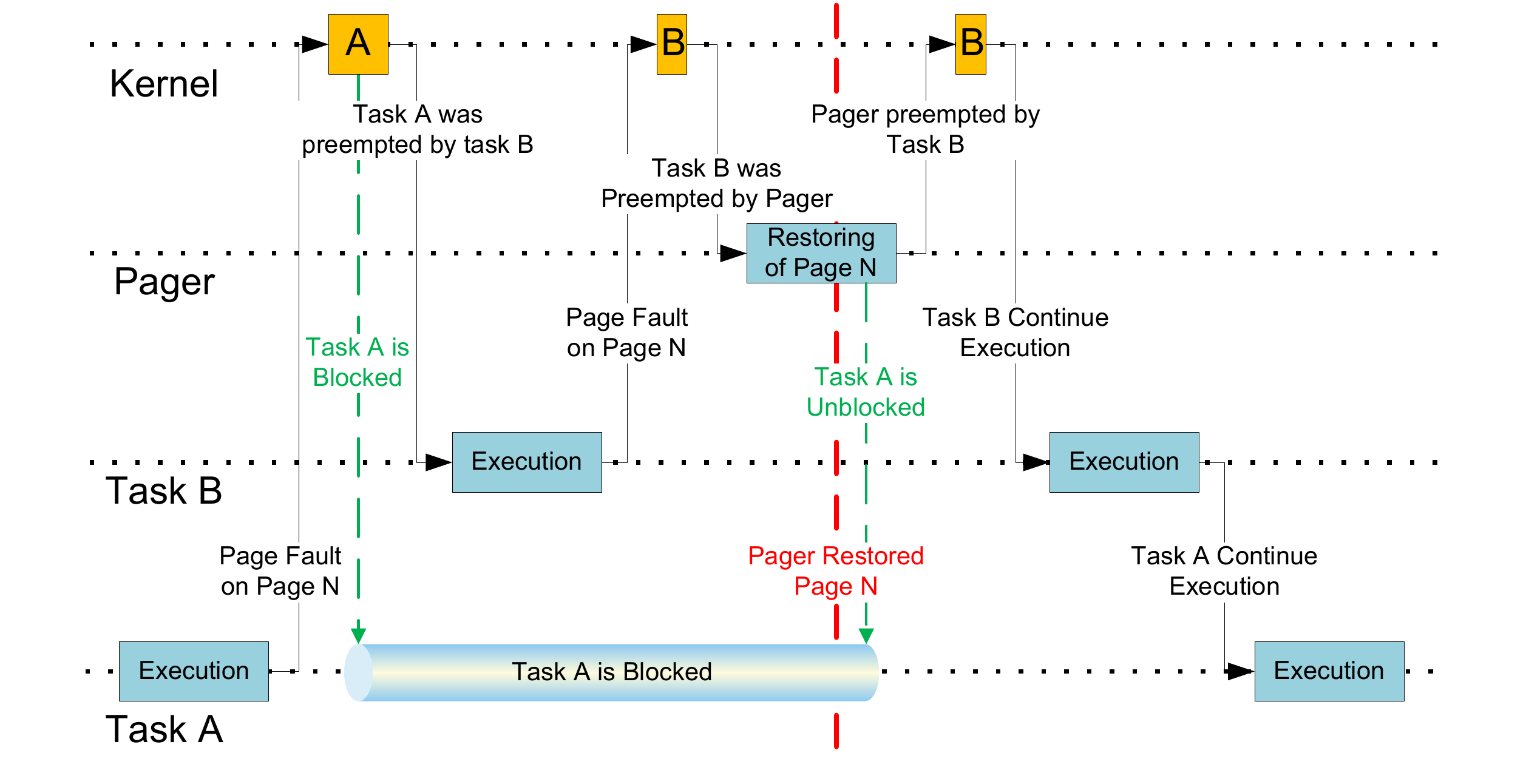} \hfil 
		\caption{Concurrent Page Fault Handling Case}
		\label{fig6}

		\vspace{0.5cm}
		
		\hfil \includegraphics[width=15cm]{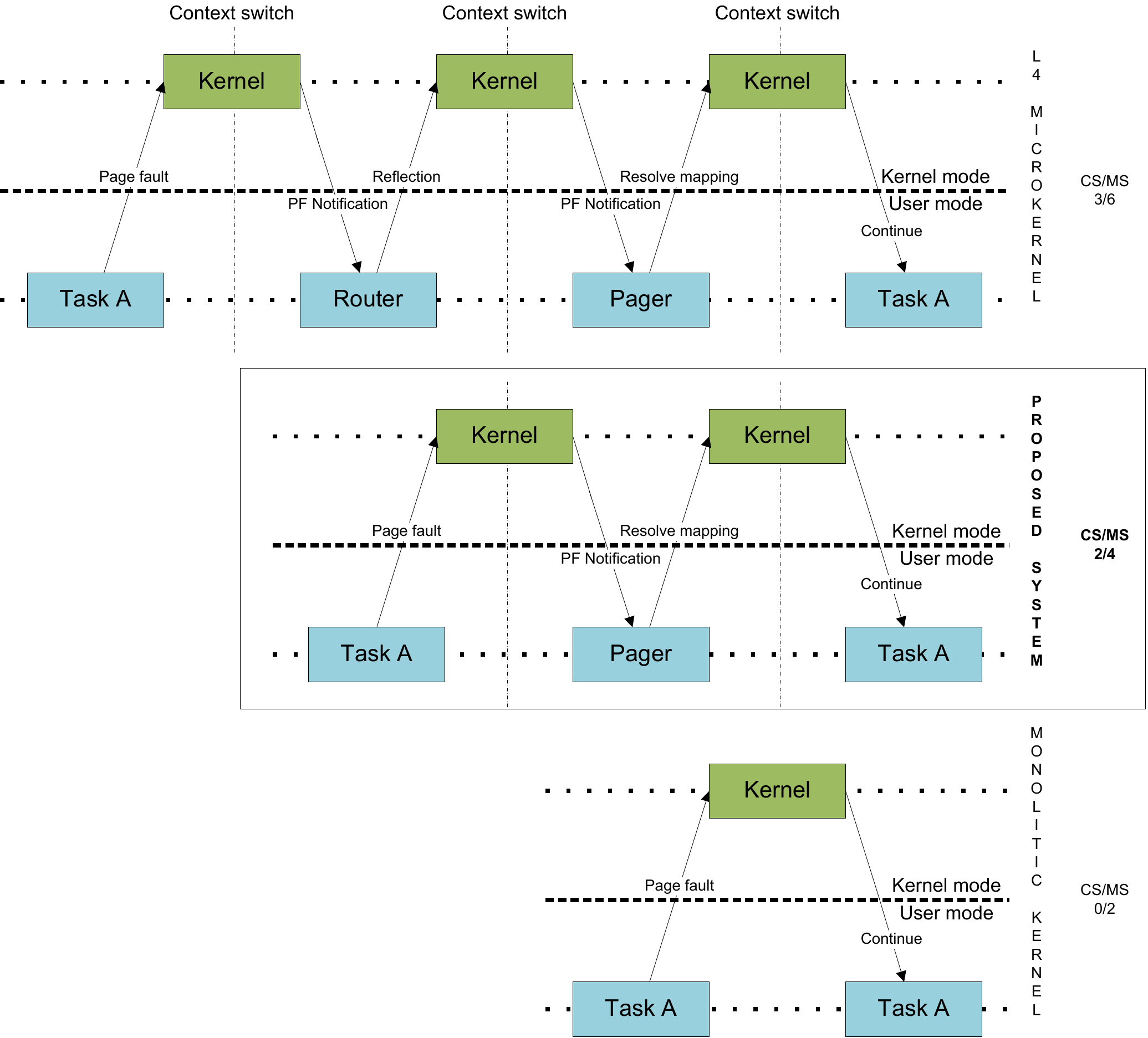} \hfil 
		\caption{Page Fault Handling Schemes}
		\label{fig7}

		\vfill
	\end{figure*}

	Let's consider the case illustrated on Fig. \ref{fig6}. 
	In this case, there are two threads A and B concurrently running in the same virtual address space. 
	Each of them made page fault on the same page N sequentially one after another. 
	But thread A and thread B achieved different handling from the kernel side. 
	Kernel notified pager that thread A triggered page fault on page N via message and blocks thread A execution until pager will have page N restored.
	Then when pager got a CPU time it restored mapping of the page N and unblocked thread A, allowing it future execution.

	The case of thread B differs from the case of thread A in the fact, that between actual page fault generation and the end of first phase of page fault handling, pager has scheduled for CPU time and already restored mapping of the page N.
	As a result, kernel can simply and safely return control back to the thread B without its blocking/unblocking and additional pager involving. 

	\section{CASE STUDY, PRELIMINARY EVALUATION AND DISCUSSION}

	Lets consider three page fault handling schemes used in multi-pager environment supporting operating systems: monolithic kernel approach, Fiasco.OC approach and finally proposed approach (Fig. \ref{fig7}).
	For each case we consider a user mode thread that triggers a page fault exception which must be handled by paging server to allow the faulted thread to continue its execution.
	We consider a general path of the page fault handling without taking into the account the performance penalty introduced by paging server, cost of transition between kernel mode and user mode and cost of IPC.

	Originally, the first proposed page fault handling cycle was proposed for  monolithic kernel design. 
	According to this approach all the page fault handling activities are performed in the kernel. 
	Kernel is the only paging server of the system and can be considered as a tightly integrated set of subsystems, which includes multiple paging modules.
	Due to this we can consider monolithic kernel as a multi-pager environment provider.
	Page fault handling cycle in the environment of monolithic operating system includes two crossing of the kernel mode/user mode boundary: one is triggered by page fault exception and the second one is to return the control flow to the faulted thread.
	All page fault resolution actions are performed in the kernel without involving another threads.
	This page fault handling scheme is the most efficient, but not applicable for the second generation microkernel design.
 
	The key design principle used in the first second generation microkernel L4 was a minimality principle, in accordance to which as much as possible functionality was pushed out from the kernel.
	Page fault handling  is a part of functionality that was removed from the kernel.
	Instead of full-featured page fault handling kernel only dispatches page fault exceptions generated by CPU through message passing to the dedicated page fault handling server thread.
	This thread called pager is explicitly assigned to the threads running in the system on the one pager per thread basis.
	As a result, pager is wired not to virtual address space but to thread and it is responsible for handling all page faults generated by the thread to which it was assigned.
	Page fault handling cycle of the L4 is similar to the same in Mach and includes four crossings of the kernel mode/user mode boundaries and two thread context switches.
	But as you can see there is no multi-pager environment support implemented in the kernel.

	The researchers from TUD noted multi-paging importance and proposed to implement its support on the level of runtime.
	They implemented this support in L4Re runtime, which creates a special pager thread per process and assigns it as a pager for each thread running in the same process.
	This pager thread maintains the database of mappings between virtual address space regions and pagers assigned to it.
	Page fault handling cycle in this case includes 6 crossings of kernel mode/user mode transitions and 3 thread context switches.
	At the first step CPU switches from user to kernel mode in reply to page fault exception triggered by running thread A.
	Kernel in reply suspends thread A and sends page fault message to the L4Re pager task assigned to task A. 
	Pager in its turn consults with mapping database and figures out which task is responsible for resolution of page fault. 
	Pager resends (reflects) the initial page fault message to the actual pager identified on the previous step.
	After this the actual pager finally can perform the actions for actual servicing of the page fault of task A.
	
	\begin{table}[!h]
		\renewcommand{\arraystretch}{1.2}
		\centering

		\caption{COMPARISON OF PAGE FAULT HANDLING IN DIFFERENT MULTI-PAGER ENVIRONMENTS} 
		\label{table1}

		\begin{small}
			\begin{tabular}{ | >{\centering\arraybackslash}m{2.5cm} | >{\centering\arraybackslash}m{2cm} | >{\centering\arraybackslash}m{2cm} |  }
				\hline
				{\bfseries Architecture} 						& {\bfseries Mode switch count} 	& {\bfseries Context switch count}  	\\ \hline 
				\raggedright Monolithic kernel					& 2					& 0						\\ \hline
				\raggedright Proposed approach Single-paging L4 Mach	& 4					& 2						\\ \hline
				\raggedright L4 Microkernel + L4Re (Fiasco.OC)		& 6					& 3						\\ \hline
			\end{tabular}
		\end{small} 
	\end{table}

	Proposed approach preserves the same page fault handling cycle as an original L4 kernel, but with natural intra-kernel support of multi-pager environment. 
	On the other hand it is similar to the simplified  page fault handling scheme of the Mach microkernel acceptable for the second generation microkernels.
	Page fault triggered by the thread causes transition from the user mode to kernel mode where kernel at the final step of its dispatching identifies the pager thread assigned to the region of address space which the faulted address belongs to.
	Then similar to other exception handling it suspends the faulted thread and sends page fault description message to the pager identified.
	After resolution of the page fault, pager notifies the kernel about resolution results and by this resumes faulted thread.
	Page fault handling cycle is accomplished by passing control back to the resumed thread.
	As a result four transitions between kernel and user mode and two thread context switches are required by proposed approach for page fault handling cycle. Results are summarized in table \ref{table1}.
	
	Proposed approach allows to reduce overhead of the page fault handling in terms of mode switch and thread context switch by 33.3\% while preserving multi-pager environment support.
	In the same time it introduces only a very little additional code complexity and incurs only 4Kb of memory overhead per virtual address space.
	But note that L4 + L4Re approach preserves similar per address space memory overhead but on the runtime level, because code of the L4Re task and mapping database is enforced to be located on the unswappable memory, as it must eliminate page faults which can be triggered by L4Re pager thread itself.

	In fact the proposed approach can be criticized from the point of view of minimality principle. 
	But it can be advocated by the same arguments which was used for the advocation of the intra-kernel scheduling. 
	Indeed, microkernel looks like a natural location for the multi-pager environment support.
	Additional code complexity is negligible and can be measured by only a few hundreds bytes of code. 
	Unfortunately we can't provide an exact number of additional microkernel footprint bytes because the prototype of the proposed approach has been implemented as a part of written from the scratch kernel instead of changing the original L4 microkernel.
	But note also that despite the fact that it introduces memory overhead by one memory page per virtual address space the overall memory overhead of the system is reduced. 
	In contrast to the L4Re approach, there is no requirements for additional region mapper task per virtual address space and resources used by it. 

	Cost of the transition between kernel and user modes, intra-kernel exception dispatching, thread context switch and IPC are main contributors to the page fault handling overhead.
	Proposed approach adds only negligible overhead in less than dozen of simple processor instructions to the original single-pager L4 page fault handling cycle critical path.
	This additional overhead is much smaller than the cost of the transition between kernel/user modes or thread context switches which usually takes more then hundred processor cycles.
	Unfortunately fair comparison of the page fault handling cost of original L4, Fiasco.OC and proposed approach is hard to take in our current environment, because in contrast to the L4 microkernels our research kernel relies on the asynchronous IPC (reasons behind this design choice are out of the scope of this paper).
	But we believe that the analytical comparison and discussion of the proposed approach outlined in this paper is clear and sufficient to highlight benefits of the proposed solution.

	In general by this paper we wanted to advocate return of the multi-pager environment support into  kernel.
	This can be considered as a step back to the Mach design, but with preserving the general second generation design principles and choices, and with entire simplification of the mechanisms used in accordance with minimality principle.

\section{CONCLUSION AND FUTURE WORKS}
	The new way of the multi-pager environment support in the context of second generation microkernel based operating systems is proposed and described.
	It is showed that the demonstrated mechanism is superior because it introduces less overhead through reduction of number of mode switches and thread context switches performed during page fault handling cycle, provides more simple design and more flexible and natural environment for the system building.
	Despite the fact that this way of multi-pager support introduces some additional code complexity, this complexity is very small and can be advocated by the same arguments used for kernel-level scheduling advocation.
	Future work can be done on the base of this approach to explore other aspects of memory management in the context of second generation microkernels and designs of the full-featured multi-pager environment that can be built in the user mode using the proposed multi-pager support in microkernel. 
\bibliographystyle{IEEEtran}

\bibliography{IEEEabrv,IEEEexample}

\enlargethispage{-00mm}

\end{document}